\documentclass[10pt,journal,compsoc]{IEEEtran}

\usepackage{amssymb}
\usepackage{amsmath}
\usepackage[ruled,vlined]{algorithm2e}
\usepackage{array}
\usepackage{color,xcolor}
\definecolor{light-gray}{gray}{0.95}    
\definecolor{orange}{rgb}{1,0.5,0}      
\definecolor{orange}{RGB}{255,127,0}    
\definecolor{orange}{HTML}{FF7F00}      
\definecolor{orange}{cmyk}{0,0.5,1,0}   
\ifCLASSOPTIONcompsoc
  \usepackage[nocompress]{cite}
  \usepackage[caption=false,font=footnotesize,labelfont=sf,textfont=sf]{subfig}
\else
  \usepackage{cite}
  \usepackage[caption=false,font=footnotesize]{subfig}
\fi
\usepackage{dblfloatfix} 

\usepackage{epstopdf}
\usepackage{graphicx}
\usepackage{hyperref}
\usepackage[utf8]{inputenc}
\usepackage{longtable}
\usepackage{makecell}
\usepackage{mathrsfs}
\usepackage{multicol}
\usepackage{multirow}
\usepackage{textcomp}
\usepackage{url} 

\mathchardef\mhyphen="2D
\newcommand{\PreserveBackslash}[1]{\let\temp=\\#1\let\\=\temp}          
\newcolumntype{C}[1]{>{\PreserveBackslash\centering}p{#1}}              
\newcolumntype{R}[1]{>{\PreserveBackslash\raggedleft}p{#1}}
\newcolumntype{L}[1]{>{\PreserveBackslash\raggedright}p{#1}}            

\begin{document}

\title{Active CT Reconstruction \\with a Learned Sampling Policy}

\author{Ce~Wang,~\IEEEmembership{Member,~IEEE,}
       Kun~Shang,~
       Haimiao~Zhang,~\IEEEmembership{Member,~IEEE,}
       Shang~Zhao,~
       Dong~Liang,~\IEEEmembership{Senior Member, IEEE}
      and~S.~Kevin~Zhou,~\IEEEmembership{Fellow,~IEEE}

\IEEEcompsocitemizethanks{
\IEEEcompsocthanksitem C. Wang is with the Key Lab of Intelligent Information Processing of Chinese Academy of Sciences (CAS), Institute of Computing Technology Chinese Academy of Sciences, Beijing, CN.
\IEEEcompsocthanksitem K. Shang and D. Liang are with the Research Center for Medical AI, Shenzhen Institutes of Advanced Technology, Chinese Academy of Sciences, Shenzhen, Guangdong CN. D. Liang is also affiliated with Pazhou Lab, Guangzhou, CN.
\IEEEcompsocthanksitem H. M. Zhang is with the Institute of Applied Mathematics, Beijing Information Science and Technology University, Beijing, China.
\IEEEcompsocthanksitem S. Zhao and S.~Kevin~Zhou are with the School of Biomedical Engineering \& Suzhou Institute for Advanced Research,
Center for Medical Imaging, Robotics, Analytic Computing \& Learning (MIRACLE), 
University of Science and Technology of China, Suzhou, CN. Zhou is also affiliated with the Key Lab of Intelligent Information Processing of Chinese Academy of Sciences (CAS), Institute of Computing Technology, CAS, Beijing, CN.
\IEEEcompsocthanksitem C. Wang and K. Shang contributed equally to this work.
\IEEEcompsocthanksitem S. Kevin Zhou is corresponding author (skevinzhou@ustc.edu.cn).
}
}

\markboth{Journal of \LaTeX\ Class Files,~Vol.~XX, No.~X, XX~XX}%
{Shell \MakeLowercase{\textit{et al.}}: Bare Demo of IEEEtran.cls for Computer Society Journals}
%



\IEEEtitleabstractindextext{
\begin{abstract}
Computed tomography (CT) is a widely-used imaging technology that assists clinical decision-making with high-quality human body representations. To reduce the radiation dose posed by CT, sparse-view and limited-angle CT are developed with preserved image quality. However, these methods are still stuck with a fixed or uniform sampling strategy, which inhibits the possibility of acquiring a better image with an even reduced dose. In this paper, we explore this possibility via learning an active sampling policy that optimizes the sampling positions for patient-specific, high-quality reconstruction. To this end, we design an \textit{intelligent agent} for active recommendation of sampling positions based on on-the-fly reconstruction with obtained sinograms in a progressive fashion. With such a design, we achieve better performances on the NIH-AAPM dataset over popular uniform sampling, especially when the number of views is small. Finally, such a design also enables RoI-aware reconstruction with improved reconstruction quality within regions of interest (RoI's) that are clinically important. Experiments on the VerSe dataset demonstrate this ability of our sampling policy, which is difficult to achieve based on uniform sampling.
\end{abstract}

}

\maketitle

\IEEEdisplaynontitleabstractindextext

%
\IEEEpeerreviewmaketitle

\IEEEraisesectionheading{\section{Introduction}\label{sec:introduction}}

%
%
%
%

\IEEEPARstart{C}{omputed} tomography (CT), a widely-used imaging technology, is able to reconstruct highly-detailed,  cross-sectional maps of an object. Clinical diagnosis depends on the quality of the CT image, which shows the interior composition of the object. Note that the technology works by measuring the intensity decay of photons emitted from an X-Ray tube along a set of predefined rays that traverse the object. \textit{Normally, the higher the CT dose, the clearer the image}. Unfortunately, the carcinogenic nature of ionizing radiation means one needs to reduce the radiation dose that the patient is exposed to in a CT scan, which raises safety concerns for patients. In contrast, the lower dose means noisier data and, therefore, a more challenging imaging problem. To balance the demand of two aspects, sparse-view (SV) and limited-angle (LA) imaging technologies are developed \cite{Wang2020Deep,zhou2019handbook}, which acquire sinograms using a limited number of angles and within a limited angle range, respectively. 

Traditional methods that reconstruct an image from limited acquisition focus on iterative methods with representative knowledge-based priors, such as Total Variation based methods~\cite{mahmood2018adaptive, sidky2008image}, Non-local based method~\cite{zeng2015spectral}, Sparsity-based method~\cite{bao2019convolutional}, and Low-Rank-based methods~\cite{kim2014sparse, semerci2014tensor, gao2011robust}. However, these knowledge-based approaches typically introduce an iterative reconstruction process that is computationally expensive. Recently, the deep learning (DL) based approaches, such as FBPConvNet~\cite{jin2017deep} and CasRedSCAN~\cite{zhou2021limited}, have been proposed and achieved satisfactory performances with less time consumption, which directly learn a mapping function between the low-quality FBP reconstruction and clean high-quality images via paired training. Such a training procedure introduces a risk of falling into the local minima of training distribution, thereby unable to generalize to unseen signals, especially when there is a low number of training images. Compared to natural images, medical images (\textit{e.g.} CT images) are sparser. To improve the generalizablity, a series of deep-unfolding frameworks are proposed~\cite{sun2016deep, yang2018admm, adler2018learned, wang2021improving, zhang2020metainv, chen2018learn, wu2021drone,cheng2020learned} by further combining the iterative algorithms with DL, which inherits both the generalizability of knowledge-based methods and the high-performances of DL-based methods and hence boosts the reconstruction performances of SV/LA CT by a large margin. 

Despite the success achieved by the above methods, CT imaging technology still undergoes development. Individual factors, such as weight, age, and sex, make body representation different. Such differences are not considered in the uniform trajectory, making the sampling \underline{non-adaptive} to patients~\cite{shen2020learning}. Nevertheless, few works study the sinogram sampling process, which seems helpful in MRI reconstruction~\cite{zhang2019reducing, pineda2020active,jin2019self}. UF-AEC~\cite{shen2020learning} marks the first attempt that simultaneously learns the trajectory with PD-Net~\cite{adler2018learned} in CT imaging. They simultaneously optimize the radiation dose of each X-Ray signal and the sampling trajectory, and return higher rewards when the selected sinogram brings the most improved peak signal-to-noise ratio (PSNR). While such a design simultaneously learns the trajectory and tunes the dose distributed along with each X-Ray signal, we hypothesize that further by taking into account patient-specific diagnosis requirements, CT reconstruction should provide better visual quality in corresponding clinically concerned regions. For instance, clinical diagnosis of patients with spinal diseases needs better image quality in the spine region; whereas COVID-19 examination necessitates more details in lung reconstruction. Although the commonly-used uniform trajectory provides overall better reconstructions, they fail to present the required image quality granularity in Region of Interest (RoI) across different patients or from specific diagnosis requirements, making the uniform trajectory \underline{clinically ineffective} for personalized clinical decision making. Note that a physically developed method~\cite{chityala2004region,heuscher2011ct} for CT imaging is using a controllable beam filter to impose a high radiation dose only in the RoI, resulting in a better RoI reconstruction. However, the dose distributed inside and outside RoI needs to be carefully balanced for different demands. Till now, to the best of our knowledge, there are few works considering this problem from an algorithmic perspective.

In this paper, we incorporate active sinogram sampling into our reconstruction framework and learn to jointly acquire projections and reconstruct the image, which adapts to the patient-specific diagnosis requirements. Specifically, we separately learn the two subtasks with two corresponding modules, one for learning the sampling trajectory, named Intelligent Agent (\textbf{A}), and the other for the reconstruction, named Reconstructor (\textbf{R}). Facing no ground truth for \textbf{A}, we pretrain the module \textbf{A} to be capable of ranking sinograms in terms of the contribution to CT reconstructions via metric learning. Then, the pretrained module \textbf{A} is incorporated to recommend the most reconstruction-benefiting sinogram sampling position. Next, we design an iterative reconstruction process to gradually (i) rank all projected sinograms of the current reconstruction with \textbf{A} and select the candidates with higher scores; and (ii) reconstruct with sinograms composed of the previous stages and the selected candidates in (i). To compensate for the optimization shift between learning better ranking with \textbf{A} and better reconstruction with \textbf{R}, we employ an alternating optimization design. Via sufficient training, our method learns to actively obtain sinograms benefiting high-quality reconstructions and is able to go beyond what can be done by doing either separately.

In sum, our contributions are as follows:
\begin{itemize}
	\item Towards the optimal sampling for an active reconstruction, we design an iterative reconstruction process between the \textbf{R} and \textbf{A} modules, one for reconstructing images and the other for recommending the sampling trajectory, respectively. To alleviate the different optimization directions of these two modules, we propose an alternative pipeline to gradually adopt the actively learned 
    sampling trajectory in training.
	\item In order to make the learned sampling trajectory controllable and matching with real scenarios, we introduce multiple hyperparameters in our iterative reconstruction. For allowing more variations in training, we relax the searching space of hyperparameters. Then, taking into account the difference across patients and different clinical requirements, we tighten the space during inference for better performances.
	\item Empirical experiments on NIH-AAPM~\cite{mccollough2016tu} and VerSe~\cite{SEKUBOYINA2021102166,deng2021ctspine1k} benchmark datasets demonstrate that our proposed learning-based policy achieves better reconstruction on both the whole image and the predefined RoI. The visualization of learned sampling mask verifies the adaptiveness and its effectiveness on the RoI information.
\end{itemize}

\section{Motivation}

CT reconstruction aims to reconstruct structural representations of external/internal tissues of the human body $u \in \mathbb{R}^{HW}$ ($H$ and $W$ are the image height and width) from the corresponding sinogram ${y}_{f} \in \mathbb{R}^{TD}$ ($T$ is the sampling times, and $D$ is the number of detector photons), which is formulated as
\begin{equation*}
{y}_{f}= A_f u + n, 
\end{equation*}
where the fully-forward projection matrix $A_f \in \mathbb{R}^{TD \times HW}$ represents the full-sampling Radon transform and $n$ denotes the imaging photon noise. In SV/LA CT, the sampling times $M$ is much less than $T$, such that we define an additional sub-sampling matrix $P\in{{\mathbb{R}}^{MD \times TD}}$, composed of \{0, 1\} elements to choose suitable subvectors from ${y}_{f}$. Thus we have the down-sampled observation $y= P{y}_{f}$.


In iterative methods, DL methods, or  deep-unrolling methods, the sampling matrix $P$ is pre-defined and the included trajectory is uniformly distributed around the patient (usually limited in 180$^{\circ}$ since the penetrability of X-Ray leads to repetitive information). This evenly-acquired sinogram collects global body representation and brings overall satisfactory reconstruction.
\begin{figure}[h!]
\begin{center}
\includegraphics[width=0.5\textwidth]{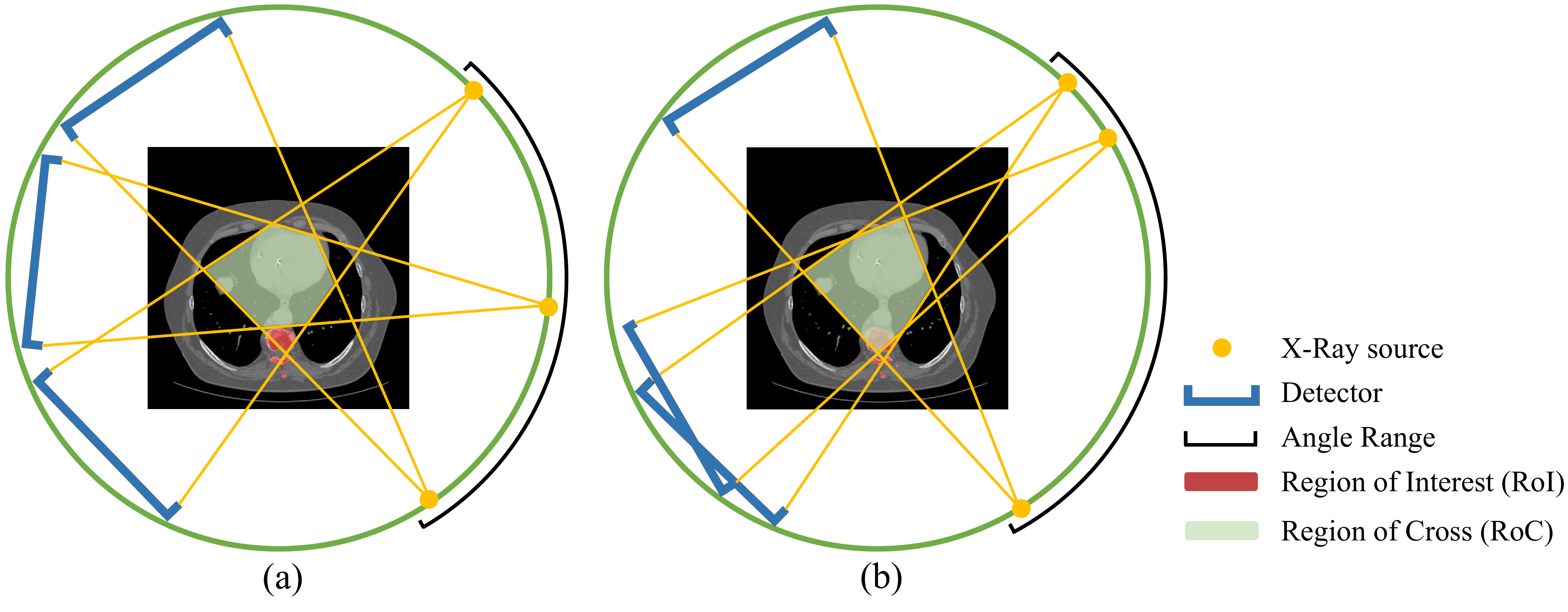}
\end{center}
\caption{We visualize a CT image with a Region of Interest (RoI) labeled in red that covers the spine, say for diagnosing spinal diseases. With the commonly-used uniform sampling, the Region of Cross (RoC) by acquired sinograms mismatches with RoI as in (a), which is not clinically expected. Then, with rearrangement of sampling trajectories as in (b), the RoC overlaps better with RoI, leading to a possibly better RoI reconstruction.}
\label{motivation}
\end{figure}
However, such predefined sinogram sampling trajectories do not consider patient-specific or task-specific characteristics. Taking spine checking as a motivating example in Fig.~\ref{motivation}, we here only consider three partial sinograms within a fixed angle range (labeled with curves) to simplify the problem. With the uniform trajectory (three sinograms distributed equally within the fixed angle range) as in Fig.~\ref{motivation} (a), the Region of Cross (RoC) formed by acquired sinograms does not bear a significant overlap with the RoI. Indeed, this is not expected for the shown CT image with the RoI (spine parts) deviating a lot from the center, resulting in the reconstruction that seems overall clear but with degraded quality in the RoI. The phenomena makes the deployed uniform sampling trajectories less effective when clinicians requires to examine the patient's spinal region. \textit{To avoid such undesirable properties in uniform trajectory, additional optimization of the sampling matrix $P$ results in a dynamic arrangement of sampling trajectory, providing a potential to achieve a personalized and clinically effective imaging process.} 

\begin{figure}[h!]
	\includegraphics[width=0.5\textwidth]{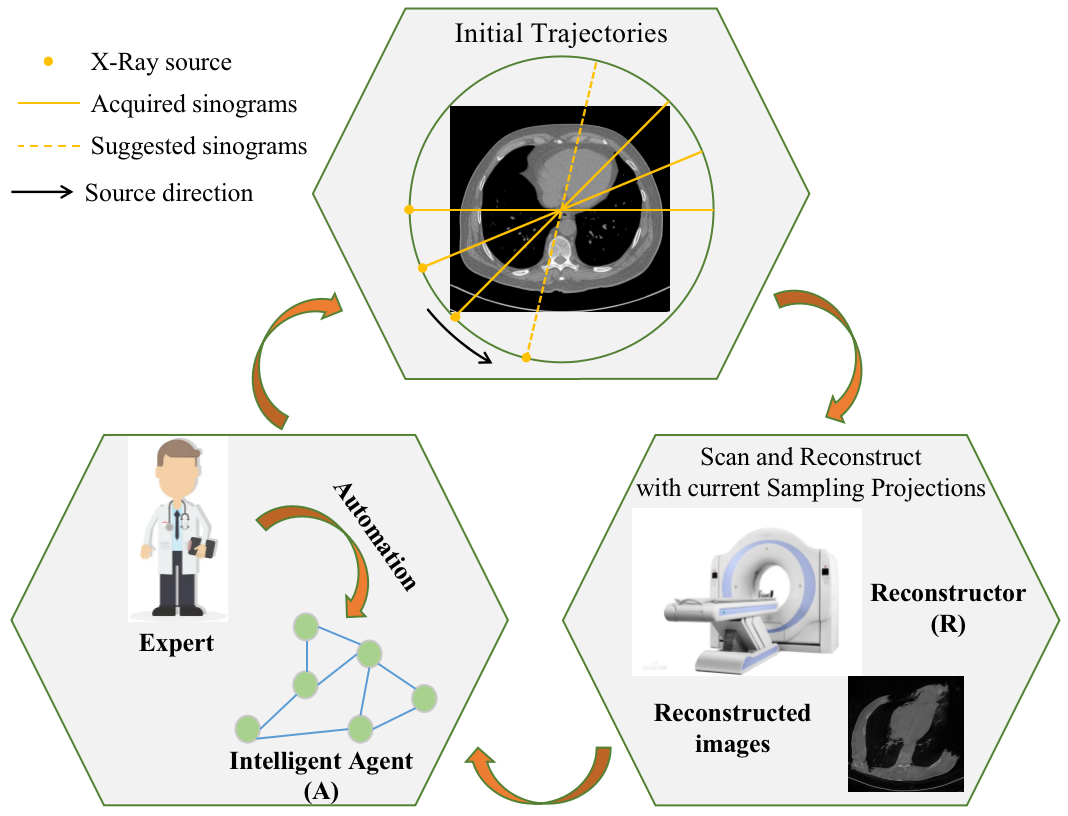}
	\caption{The iterative process of simultaneous sampling and reconstruction.}
	\label{intro}
\end{figure}

Whereas, with a rearrangement of the sampling positions as in Fig.~\ref{motivation} (b), the resulted RoC highly overlaps with the RoI. Motivated by the insignificant overlap between RoC and RoI in uniform sampling and the RoC changes with rearrangement, we explore the option of designing \textbf{an active sampling policy} to gradually learn the sampling matrix $P$ and reconstruction simultaneously in an iterative process as the workflow depicted in Fig.~\ref{intro}. 

Till now, only Shen~\textit{et~al.}~\cite{shen2020learning} use reinforcement learning to design a novel dynamically sampling policy based on the assumption of a random access to any sampling angle at any time, which is therefore too radical for practical deployment since dynamically tuning the radiation dose and sampling positions is physically impractical in CT machine per their discussion. Note that \textit{both the sinogram trajectory and the tuned radiation dose gradually contribute to the reconstruction performance until the total number of sinograms reaches the pre-defined upper bound.} Besides, considering each selected sinogram, dynamically changing the suitable position and tuning the accompanying radiation dose both change the accumulated dose distribution around the patient. 

As technology advances, dynamically moving the radioactive source is realizable. Thus, further optimizing $P$ for a better reconstruction has recently become attractive yet still challenging. Targeting this issue, in this paper, we present a design with a controllable sampling trajectory, and learn the optimal matrix $P$ simultaneously with reconstruction to impose the positive interaction, which better aligns with real scenarios.

\begin{figure}[!t]
	\includegraphics[width=0.45\textwidth]{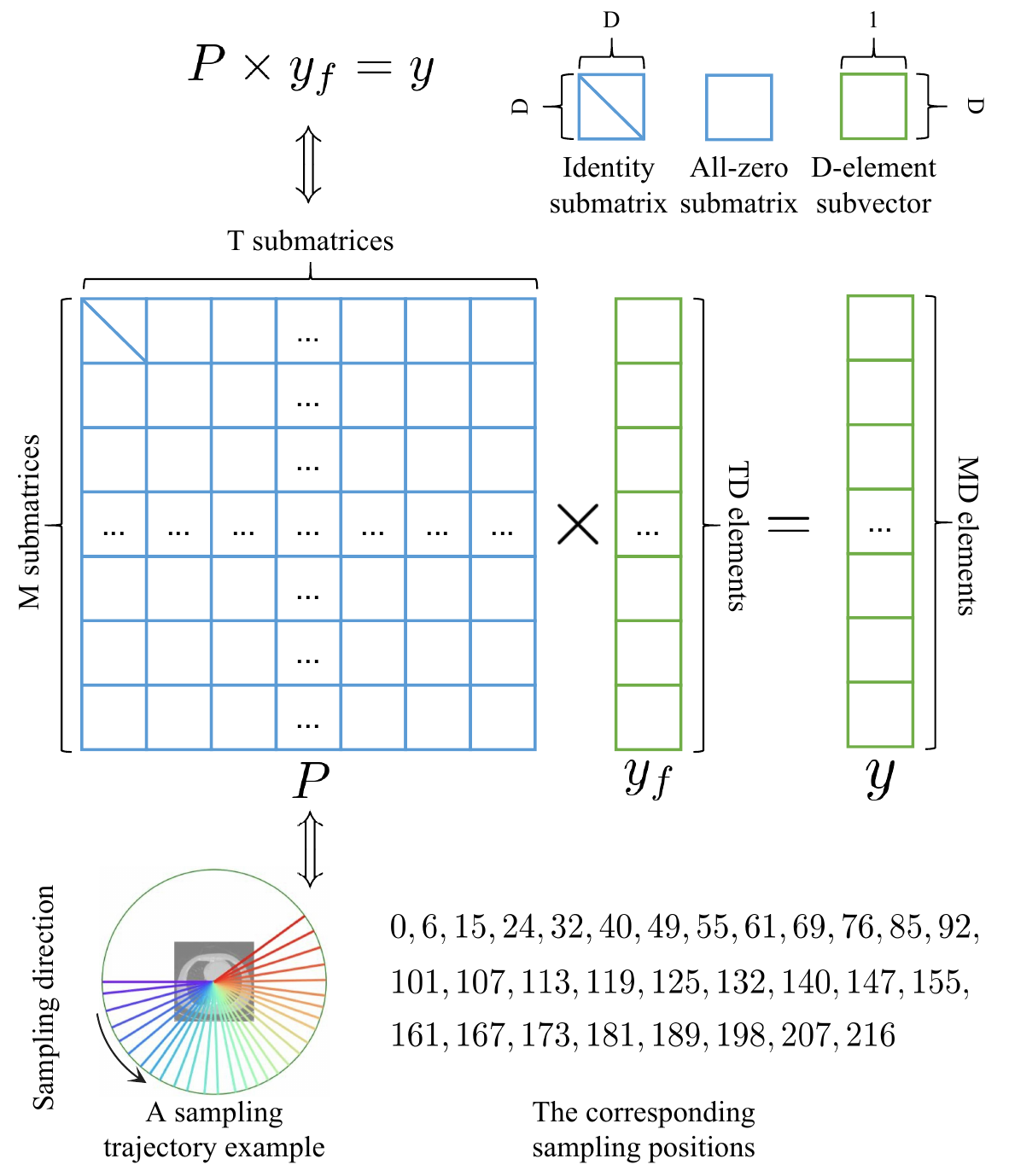}
	\caption{We visualize the down-sampling process of an example, where we represent $P$ with a block matrix with identity submatrices and all-zero submatrices. The optimization of P is equivalent to selecting the sampling positions for identity submatrices.}
	\label{mask_p}
\end{figure}

\begin{figure*}[!t]
	\includegraphics[width=\textwidth]{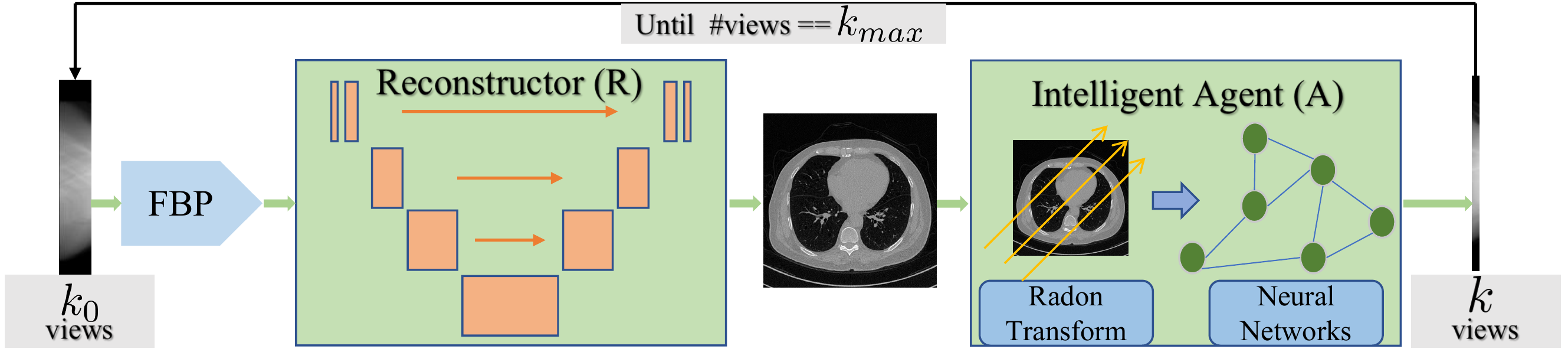}
	\caption{The framework of our active sampling and reconstruction process. With initialized ${k}_{0}$ sinograms, denoted as ${y}_{0}$, we then use a reconstructor (\textbf{R}) (realized with U-Net here) to reconstruct the corresponding CT image. The following intelligent agent (\textbf{A}) recommends the next $k$ sinogram acquisition positions, which are optimal for the currently observed reconstruction. The iterative process continues until ${k}_{max}$ sinograms are sampled. }
	\label{active-framework}
\end{figure*}

\section{Methodology}
To this end, we reformulate the reconstruction problem as follows: 
\begin{equation}
	\label{objective}
	\min_{u, P} \quad \frac{1}{2}\| P{A}_{f}~u-y\|^{2} + {R}_{1}(u) + {R}_{2}(P),
\end{equation}
where ${R}_{1}(u)$ and ${R}_{2}(P)$ are the regularization terms used to impose the prior information. Traditionally, the minimization problem (\ref{objective}) can be efficiently solved via the splitting method. The variables $u$ and $P$ in the above are split into two blocks:

\begin{equation}
	\label{iterative-objective}
	\left\{ 
	\begin{aligned}
		{P}^{k+1}&=\arg \min_{P} \frac{1}{2}\| {P}^{k}{A}_{f}~{u}^{k}-y\|^{2} + {R}_{2}(P), \\
		{u}^{k+1}&=\arg \min_{u} \frac{1}{2}\| {P}^{k+1}{A}_{f}~{u}^{k}-y\|^{2} +  {R}_{1}(u). \\
	\end{aligned}
	\right.
\end{equation}

Rethinking the above-introduced $P$ as in Fig.~\ref{mask_p}, $P$ is a very high-dimensional and sparse matrix. More specifically, it is composed of $M\times T$ submatrices, which are either an identity matrix or an all-zero matrix. Due to its special form, direct optimization of $P$ in (\ref{iterative-objective}) is difficult and computationally expensive. To speed up the iterative reconstruction, recall the physical significance of $P$. For the fixed fully sampled sinogram ${y}_{f}\in{{\mathbb{R}}^{TD}}$, $P$ adopts $M$-times (with $M\ll T$) sampling on the temporal dimension, which is represented by the identity matrix in Fig.~\ref{mask_p}, and each time a projection ${y}_{f,k}\in{{\mathbb{R}}^{D}} (1\leq k\leq T)$ is returned. In practice, ${y}_{f,k}$ represents one imaging line on the detector (\textit{i.e.}, the D-element subvector in Fig.~\ref{mask_p}). With such analysis, we simplify the optimization of $P$ to the \textbf{iterative selection} of $M$-times choice of identity matrix in our framework, that is, adding one more imaging line per iteration. In other words, it becomes a \underline{discrete selection} rather than a \underline{continuous optimization} problem.

For this purpose, we use the neural networks to learn the solutions of Eq. (\ref{iterative-objective}) in each iteration, which gives
\begin{equation}
	\left\{
	\begin{aligned}
		{P}^{k+1}& = {\textbf{A}}_{\psi}({u}^{k}, {P}^{k}), \\
		{u}^{k+1}&=  {u}^{k} + {\textbf{R}}_{\theta}( {u}^{k}, {P}^{k+1} ), \\
	\end{aligned}
	\right.
\end{equation}
where ${\textbf{R}}_{\theta}$ and ${\textbf{A}}_{\psi}$ are parameterized sub-networks to iteratively compute the intermediate estimations.  In this way, we have decoupled the optimization problem (\ref{objective}) into the acquisition step and reconstruction step, where the alternating direction method is employed to solve a multi-variable optimization problem.
The overall active acquisition$\leftrightarrow$reconstruction process is depicted in Fig.~\ref{active-framework}. The complete process is alternatively between the optimization of Reconstructor (\textbf{R}) and Intelligent Agent (\textbf{A}). Especially, we first pre-define ${k}_{0}$ initial sinogram-sampling positions (we empirically set initial positions uniformly in experiments) to give a suitable initialization of ${u}^{0}$ and ${P}^{0}$. According to the current image quality, the module \textbf{A} is then utilized to recommend the next $k$ acquisition positions within a certain angle range in the acquisition step. The proposed iterative process outputs a series of iterates:
$$ ({P}^{1}, {u}^{1}), ({P}^{2}, {u}^{2}), ({P}^{3}, {u}^{3}), \ldots $$
and continues until the total number of acquired sinograms equals ${k}_{max}$. Next, we introduce the modules \textbf{R}, \textbf{A}, and the training paradigm in detail.

\subsection{Reconstructor}
From the perspective of active learning, the obtained ${k}_{0} + nk$ sinograms after $n$-time iterations provide partial and low-quality representation, which can be utilized to suggest the next sinogram acquisition positions. To ensure the power of the latter sampling policy design, we realize \textbf{R} with the commonly-used backbone U-Net, which can also be replaced with other structures as in our later experiments. In the iterative sampling and reconstruction process, $\textbf{R}$ is used to transit estimations from different acquisition scenarios, and the following \textbf{A} would evaluate the correlation between each candidate and the final reconstruction performance. Specifically, in the $n$-th iteration, the reconstructor $\textbf{R}$ receives ${k}_{0}+nk$ acquired sinograms ${y}_{n}$, and outputs the reconstruction ${\hat{u}}_{n}$. Towards an efficient computation, we employ the same parameters among each iteration, which also renders better robustness of our reconstructor. In this way, the final loss function for $\textbf{R}$ is defined as follows:
\begin{equation}
	\label{recon-loss}
	{\cal{L}}_{\textbf{R}} = \| \hat{u} - {u}_{gt} \|^{2},
\end{equation}
where $\hat{u}$ and ${u}_{gt}$ are the final estimation and the corresponding ground-truth CT image, respectively.

\subsection{Intelligent agent}
Given an estimated reconstruction, an experienced clinician is capable of deciding which part of the reconstruction is sufficient for diagnosis, and therefore suggests a coarse sampling range to enhance the reconstruction. Nevertheless, such expert-dominated sampling suggestion is expensive because of the scarcity of clinicians. A desirable substitution is an Intelligent Agent (\textbf{A}) able to seek out the most reconstruction-benefited sampling positions with the current estimated ${\hat{u}}_{n}$. But this is an extremely difficult problem since the image domain information cannot directly guide the sinogram position searching. To solve the problem, we propose a two-stage projection-image domain correlation sampling policy, denoted by \textbf{A}, as in Fig.~\ref{active-framework}. Specifically, with the current estimation, we first transform ${\hat{u}}_{n}$ into the Radon domain with ${A}_{f}$. Obtaining the current full-projections ${A}_{f}({\hat{u}}_{n})$, we next use a fully-connected network to output a \textit{confidence score} for each sinogram, which tells the system how each obtained projection correlates with the final reconstruction. With the score, we finally select the highest $k$ candidates in a limited angle range $({\alpha}_{p}, {\alpha}_{q})$, where ${\alpha}_{p}$ and ${\alpha}_{q}$ are the pre-defined minimal and maximal distances, respectively, between these candidates and the last chosen sinogram position. The procedure provides two benefits: 
\begin{itemize}
	\item \textbf{Flexibility}. The design is flexible to be plugged in the current CT machines, just replacing the used reconstruction algorithm with ours and sample projections on the actively learned positions instead of fixed ones;
	\item \textbf{Adaptiveness}. The optimal selection for the introduced angle-range $({\alpha}_{p}, {\alpha}_{q})$ allows more freedom for the algorithm to adaptively reconstruct images according to the current image domain information of ${\hat{u}}_{n}$.
\end{itemize}

\begin{algorithm}[!t]
	\caption{Training Strategy}
	\label{alg1}
	\LinesNumbered 
	\KwIn{The ground truth CT image $u$, the involved ${\alpha}_{max}$, ${k}_{0}$, $k$, ${k}_{max}$, ${\alpha}_{p}$, ${\alpha}_{q}$, the number of epochs $T$, batch size $b=1$, learning rate $lr$, the network parameters ${\theta}$ of $\textbf{R}$, ${\psi}$ of $\textbf{A}$, and ${\beta}_{1}, {\beta}_{2}$ of Adam.}
	\KwOut{The optimized parameters $\hat{\theta}, \hat{\psi}$.}
	\For{$t=1:T$}{
		Sample initial ${k}_{0}$ sinograms ${y}_{0}$\;
		\For{$n=0:N$ ($N=({k}_{max} - {k}_{0})/{k}$)}{
			Reconstruct image $\textbf{R}_{\theta}({y}_{n})$\;
			Evaluate $\textbf{R}_{\theta}({y}_{n})$ with $\textbf{A}_{\psi}$\;
			Recommend $k$ sampling positions in $({\alpha}_{p}, {\alpha}_{q})$\;
			Sample the $k$ sinograms, denoted as the set ${\cal{K}}_{n}$\;
			Update ${y}_{n+1} = {y}_{n} \bigcup {\cal{K}}_{n}$;}
		\eIf{$t\%2=0$}{
			Update reconstructor parameters $\theta$:\
			${grad}_{\theta}=\nabla {\cal{L}}_{{\textbf{R}}_{\theta}}$\
			$\theta=\rm Adam \it ({grad}_{\theta}, \theta, lr, {\beta}_{1}, {\beta}_{2})$\
		}{
			Compute $A_{rd, i}$; \\
			Update evaluator parameters $\psi$: \
			${grad}_{\psi}=\nabla {\cal{L}}_{{\textbf{A}}_{\psi}}$\
			$\psi=\rm Adam \it ({grad}_{\psi}, \psi, lr, {\beta}_{1}, {\beta}_{2})$
		}
	}
\end{algorithm}

In such a way, the key point is to evaluate the correlation between each projection $\{ {A}_{f}({\hat{u}}_{n})_{i} \}_{i=1}^{{k}_{max}}$ of the current reconstruction and the final reconstruction, and we have designed a self-supervised strategy for the position sampling process since there are no \textit{ground-truth sampling positions for each CT image in practical reconstruction.} Firstly, we compute the reliability of the current projection position with the following metric:
\begin{equation}
	A_{rd, i} = \exp( - \|{A}_{f}({\hat{u}}_{n})_{i} - {A}_{f}({u}_{gt})_{i}  \|^{2} ),
\end{equation}
where ${A}_{f}$ is the fully-sampled transform matrix, ${\hat{u}}_{n}$ is the current reconstruction with currently-sampled sinograms, and ${u}_{gt}$ is the correspond ground-truth CT image. The metric indeed softly evaluates how the current sinograms match with ground-truth sinograms, and would assign $\rightarrow1$ to the most approximate one and $\rightarrow0$ to the fakest one. With the computed $A_{rd, i}$ as the supervision of $\textbf{A}$, our $\textbf{A}$ learns to select sinograms closest to ground-truth ones, which benefits the reconstruction mostly. The final loss function for the evaluator is as follows:
\begin{equation}
	\label{loss-IA}
	{\cal{L}}_{\textbf{A}} = \| \sum_{i} \textbf{A}(\hat{u}_{n})_{i}  - A_{rd, i} \|^{2}.
\end{equation}
The design, in fact, explores the intrinsic information from each image ${\hat{u}}_{n}$ itself. 

\subsection{Training strategy}
As described above, the included $\textbf{R}$ and $\textbf{A}$ modules target different functions: one for reconstruction and the other for selecting reliable sinograms. Simultaneously optimizing them together makes it confusing for the whole system. Motivated by the optimization procedure in GANs~\cite{goodfellow2014generative, arjovsky2017wasserstein, wang2019label}, we propose to optimize them in an alternative fashion as depicted in Algorithm~\ref{alg1} to compensate for the optimization direction shift. Concretely, we first train $\textbf{R}$ for 2 epochs with fixed $\textbf{A}$ since the training of such reconstruction is much easier than $\textbf{A}$. Then, with fixed $\textbf{R}$, we optimize $\textbf{A}$ to search for a better sampling trajectory that is most suitable for the current $\textbf{R}$. With sufficient training, the two modules would cooperate, targeting a better reconstruction.

\begin{algorithm}[t]
	\caption{Active Sampling and Reconstruction}
	\label{alg2}
	\LinesNumbered 
	\KwIn{The optimized paramenters $\hat{\theta}, \hat{\psi}$, and the involved ${\alpha}_{max}^{'}$, ${k}_{0}^{'}$, $k^{'}$, ${k}_{max}^{'}$, ${\alpha}_{p}^{'}$, ${\alpha}_{q}^{'}$.}
	\KwOut{The reconstruction $\hat{u}$.}
	Sample initial $k_{0}^{'}$ sinograms ${y}_{0}^{'}$\;
	\For{$n=0:N$ ($N=({k}_{max}^{'} - {k}_{0}^{'})/{k^{'}}$)}
	{
		Reconstruct image $\textbf{R}_{\hat{\theta}}({y}_{n}^{'})$\;
		Evaluate $\textbf{R}_{\hat{\theta}}({y}_{n}^{'})$ with $\textbf{A}_{\hat{\psi}}$\;
		Recommend $k^{'}$ sampling positions in $({\alpha}_{p}^{'}, {\alpha}_{q}^{'})$\;
		Sample the $k^{'}$ sinograms, denoted as the set ${\cal{K}}_{n}^{'}$\;
		Update ${y}_{n+1}^{'} = {y}_{n}^{'} \bigcup {\cal{K}}_{n}^{'}$;
	}
	Reconstruct the final CT image $\hat{u}$ with ${y}_{N-1}^{'}$.
\end{algorithm}

\begin{table*}[t!]
	\setlength\tabcolsep{3pt}
	\begin{center}
		\caption{Quantitative comparison of our SAS and GDS policies with RS and US. Reconstructor performance with our policy achieves significant improvement with $p<0.001$ in terms of PSNR when $15<{k}_{max}<60$. For ${k}_{max}=90$, where sampling freedom is very limited, our policy achieves significant improvement with $p < 0.05$ in terms of PSNR. The best performance in each column highlighted in \textbf{bold} and the second best is \underline{underlined}.}
		\label{active-ablation}
		\resizebox{\textwidth}{!}{
			\begin{tabular}{  l | lll | lll | lll | lll }
				\hline
				\multirow{2}{*}{{NIH-AAPM}}
				& \multicolumn{3}{c|}{{ $k_{max}$ = 15}}
				& \multicolumn{3}{c|}{{ $k_{max}$ = 30}}  &\multicolumn{3}{c|}{{$k_{max}$ = 60}}   &\multicolumn{3}{c}{{$k_{max}$ = 90}}    \\ 
				\cline{2-13}
				& PSNR(dB) & SSIM &RMSE& PSNR(dB) &SSIM&RMSE&PSNR(dB)& SSIM&RMSE&PSNR(dB)& SSIM&RMSE  \\ \hline
				RS &23.17$^{\pm1.14}$&.795$^{\pm.029}$&.070$^{\pm.010}$&25.85$^{\pm1.16}$&.836$^{\pm.015}$&.053$^{\pm.008}$&28.44$^{\pm1.10}$&.859$^{\pm.015}$&.035$^{\pm.006}$&29.40$^{\pm0.99}$&.873$^{\pm.014}$&.034$^{\pm.005}$    \\ 
				US  &24.98$^{\pm0.59}$&.813$^{\pm.017}$&.057$^{\pm.004}$&27.87$^{\pm0.59}$&.845$^{\pm.017}$&.041$^{\pm.003}$&29.75$^{\pm1.48}$ &.866$^{\pm.029}$ &.033$^{\pm.007}$&\underline{31.06}$^{\pm1.57}$&\underline{.885}$^{\pm.017}$&.029$^{\pm.006}$ \\ 
				SAS  (ours) &\underline{26.16}$^{\pm0.80}$&\underline{.828}$^{\pm.017}$&\underline{.049}$^{\pm.005}$&\textbf{28.69$^{\pm0.57}$} &\textbf{.855$^{\pm.017}$}&\textbf{.037$^{\pm.003}$}&\textbf{30.02$^{\pm0.78}$} &\textbf{.873$^{\pm.016}$}&\textbf{.032$^{\pm.003}$}&\textbf{31.28$^{\pm0.57}$}&\textbf{.888$^{\pm.016}$}&\textbf{.028$^{\pm.002}$}\\ 
				GDS (ours) &\textbf{26.59}$^{\pm0.71}$&\textbf{.829}$^{\pm.018}$&\textbf{.047}$^{\pm.004}$&\underline{28.47}$^{\pm0.66}$ &\underline{.851}$^{\pm.019}$ &\underline{.038}$^{\pm.003}$&\underline{29.84}$^{\pm0.53}$&\underline{.871}$^{\pm.017}$&\underline{.032}$^{\pm.002}$&{31.00}$^{\pm0.68}$&{.884}$^{\pm.018}$&\textbf{.028}$^{\pm.002}$\\
				\hline
			\end{tabular}
		}
	\end{center}
	\begin{center}
		\resizebox{\textwidth}{!}{
			\begin{tabular}{  l | lll | lll | lll | lll  }
				\multicolumn{7}{l}{{ Noise Level L1}}\\
				\hline 
				RS &23.30$^{\pm1.12}$&.796$^{\pm.024}$&.069$^{\pm.009}$&25.84$^{\pm1.11}$&.835$^{\pm.015}$&.052$^{\pm.007}$&28.41$^{\pm1.13}$ &.858$^{\pm.014}$ &.038$^{\pm.006}$ &29.37$^{\pm1.01}$&.872$^{\pm.014}$ &.034$^{\pm.005}$  \\ 
				US  &24.88$^{\pm0.59}$ &.809$^{\pm.017}$ &.057$^{\pm.004}$& 27.84$^{\pm0.60}$&.845$^{\pm.017}$ &.041$^{\pm.003}$&29.71$^{\pm1.48}$&.865$^{\pm.029}$ &.033$^{\pm.007}$& \underline{30.98}$^{\pm1.56}$ &\underline{.884}$^{\pm.018}$ &\underline{.029}$^{\pm.006}$   \\ 
				SAS (ours)  &\underline{26.13}$^{\pm0.75}$&\underline{.827}$^{\pm.017}$ &\underline{.050}$^{\pm.004}$&\textbf{28.65$^{\pm0.59}$} &\textbf{.854$^{\pm.017}$} &\textbf{.037$^{\pm.003}$}&\textbf{29.96$^{\pm0.78}$} &\textbf{.872$^{\pm.016}$} &\textbf{.032$^{\pm.003}$} &\textbf{31.19$^{\pm0.59}$} &\textbf{.887$^{\pm.016}$} &\textbf{.028$^{\pm.002}$}   \\ 
				GDS (ours)&\textbf{26.60$^{\pm0.73}$} &\textbf{.829$^{\pm.019}$} &\textbf{.047$^{\pm.004}$} &\underline{28.42}$^{\pm0.66}$ &\underline{.849}$^{\pm.019}$ &\underline{.038}$^{\pm.003}$ &\underline{29.78}$^{\pm0.54}$&\underline{.870}$^{\pm.018}$&\textbf{.032}$^{\pm.002}$&{30.92}$^{\pm0.67}$&{.882}$^{\pm.018}$&\underline{.029}$^{\pm.002}$\\
				\hline
			\end{tabular}
		}
	\end{center}
	\begin{center}
		\resizebox{\textwidth}{!}{
			\begin{tabular}{  l | lll | lll | lll | lll  }
				\multicolumn{7}{l}{{ Noise Level L2}}\\
				\hline 
				RS &23.39$^{\pm1.16}$&.797$^{\pm.026}$&.068$^{\pm.009}$&25.88$^{\pm1.10}$&.835$^{\pm.015}$&.051$^{\pm.007}$&28.38$^{\pm1.08}$ &.856$^{\pm.014}$ &.039$^{\pm.006}$  &29.27$^{\pm1.00}$ &.871$^{\pm.014}$ &.035$^{\pm.005}$     \\ 
				US  &24.73$^{\pm0.60}$ &.804$^{\pm.017}$ &.058$^{\pm.004}$&27.79$^{\pm0.61}$ &.844$^{\pm.017}$ &.041$^{\pm.003}$ &29.66$^{\pm1.47}$ &.863$^{\pm.029}$ &.033$^{\pm.007}$& \underline{30.88}$^{\pm1.55}$ &\underline{.882}$^{\pm.018}$ &\underline{.029}$^{\pm.006}$   \\ 
				SAS (ours)  &\underline{26.12}$^{\pm0.74}$ &\underline{.826}$^{\pm.016}$ &\underline{.050}$^{\pm.004}$ &\textbf{28.59$^{\pm0.61}$} &\textbf{.853$^{\pm.017}$} &\textbf{.037$^{\pm.003}$}&\textbf{29.91$^{\pm0.76}$} &\textbf{.870$^{\pm.016}$} &\textbf{.032$^{\pm.003}$} &\textbf{31.12$^{\pm0.62}$} &\textbf{.885$^{\pm.017}$} &\textbf{.028$^{\pm.002}$}   \\ 
				GDS (ours)&\textbf{26.58$^{\pm0.72}$} &\textbf{.828$^{\pm.018}$} &\textbf{.047$^{\pm.004}$} &\underline{28.37}$^{\pm0.65}$&\underline{.848}$^{\pm.019}$ &\underline{.038}$^{\pm.003}$&\underline{29.75}$^{\pm0.56}$&\underline{.868}$^{\pm.018}$ &\underline{.033}$^{\pm.002}$&30.82$^{\pm0.66}$&.880$^{\pm.019}$&\underline{.029}$^{\pm.002}$\\
				\hline
			\end{tabular}
		}
	\end{center}
\end{table*}

\subsection{Active sampling and reconstruction}
With the well-trained $\textbf{R}$ and $\textbf{A}$, the testing phase utilizes them to simultaneously sample a suitable trajectory and output a better reconstruction. As in Algorithm~\ref{alg2}, the hyperparameters are denoted with a superscript to distinguish them from training ones. Specifically, with pre-defined ${\alpha}_{max}^{'}$ and ${k}_{max}^{'}$ according to the physical setting, we need to additionally tune hyperparameters ${k}_{0}^{'}$, ${k}^{'}$, ${\alpha}_{p}^{'}$ and ${\alpha}_{q}^{'}$ to strive for the best one. Note that we would make these hyperparameters different in training and testing settings since (i) we release more searching freedom to the algorithm when training to help the model face with more cases; and (ii) we tighten the searching space in testing to eliminate poor ones with empirical experience. 

\subsection{RoI-aware reconstruction}
\label{roi-section}
Except for the above-mentioned benefits from the active sampling process, it also renders a potential to reconstruct CT images targeting the diagonal requirements. This is especially true for a constrained radiation budget: the Region of Cross (RoC) constructed by the limited sample positions is more likely to match with the RoI. 

Specifically, given a clinical task, the coarse Region of Interest (RoI) can be pre-defined as a 0-1 mask $\rm{M}$ similar to the segmentation masks. In such a way, we highlight the RoI reconstruction importance and Eq.~(\ref{recon-loss}) becomes:
\begin{equation}
	\label{roi-loss}
	{\cal{L}}_{\textbf{R}_{RoI}} = \| (\rm{I}+  \rm{M}) \odot (\hat{u} - {u}_{gt}) \|_{2},
\end{equation}
where \rm{I} is the identity matrix and $\odot$ is the Hadamard product. Then, with the additional guidance of $M$, the interaction between \textbf{R} and \textbf{A} results in images that pays more attention to RoI reconstruction quality with the help of rearrangement of sinogram positions.

\section{Experimental Results}
\noindent\textbf{Dataset.} The ``2016 NIH-AAPM-Mayo Clinic Low Dose CT Grand Challenge" (AAPM)~\cite{mccollough2016tu} dataset and VerSe~\cite{SEKUBOYINA2021102166,deng2021ctspine1k} dataset are included in our experiments. 
The former is used for patient-specific active reconstruction and the latter is used for RoI-aware reconstruction. For the AAPM dataset, we follow the original data partition with three anatomies, including brain, chest, and abdomen. We first conduct overall experiments on chest data 
as in Section~\ref{patient-adaptive-recon} and analyze the reconstruction stability of our method, followed by ablation experiments on the other two anatomies in Section~\ref{pathology} to verify the robustness of our method. To further demonstrate the deployment flexibility of the framework with known clinical tasks, we experiment with several VerSe volumes, which cover the whole spine of patients, to further reconstruct and optimize sampling positions for better recovery within the RoI.

\noindent\textbf{Implementation details.} Our models are implemented using the PyTorch framework. We use the Adam optimizer~\cite{kingma2014adam} with default $({\beta}_{1}, {\beta}_{2}) = (0.9, 0.999)$ to train these models. The learning rate starts from 0.0001 for the \textbf{R} and 0.0002 for the \textbf{A}. Models are all trained on a NVIDIA 3090 GPU card for 50 epochs with a batch size of 1.

\noindent\textbf{Evaluation metrics.} Reconstructed CT images are quantitatively measured by the multi-scale SSIM~\cite{wang2004image,wang2003multiscale}, PSNR, and RMSE (Root Mean Square Error), and we compute the RMSE on normalized images.

\noindent\textbf{Comparison methods.} To verify the effectiveness of our learning-based sampling process, we compare with the Uniform Sampling (US) and Random Sampling (RS). For RS, we make inference with the trained models five times and take the average to eliminate the randomness. For our method, we have proposed two policies with a different design on ${\alpha}_{p}^{'}$ and ${\alpha}_{q}^{'}$, called Sequential Active Sampling (SAS) and Global-then-Detail Sampling (GDS). The SAS keeps the angle selection range (${\alpha}_{p}^{'}$, ${\alpha}_{q}^{'}$) fixed in the whole sampling process, and the GDS firstly reconstructs the overall image with (${\alpha}_{p, g}^{'}$, ${\alpha}_{q, g}^{'}$), followed by different angle selection range (${\alpha}_{p, d}^{'}$, ${\alpha}_{q, d}^{'}$).

\begin{figure*}[t!]
	\begin{center}
		\includegraphics[width=\textwidth]{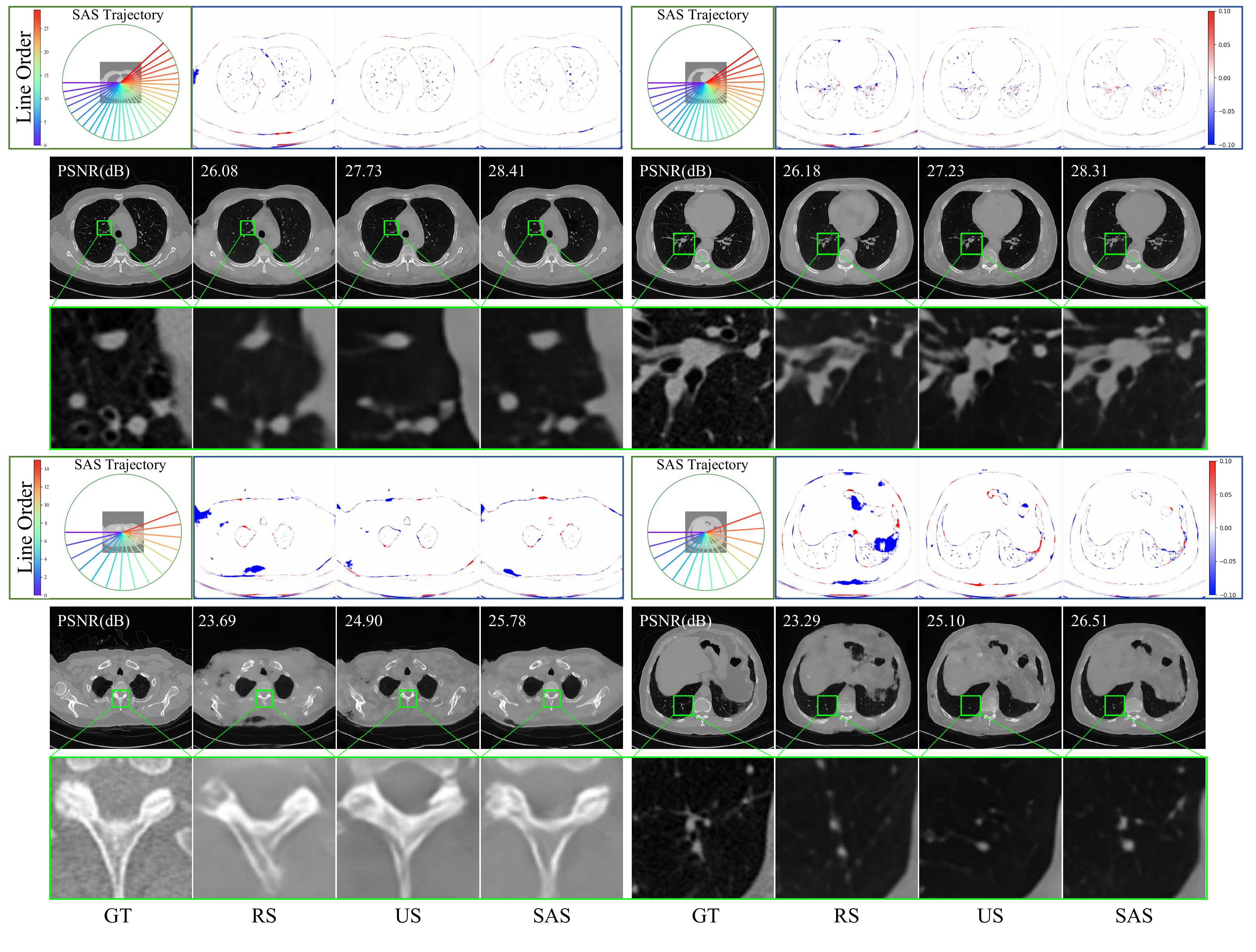}
		\caption{We visualize the reconstructions based on RS, US, and learning-based SAS in the 2$^{nd}$ and 5$^{th}$ rows, where ${k}_{max}$= 30 and 15 respectively. The corresponding difference images are shown in the 1$^{st}$ and 4$^{th}$ rows. The corresponding zoom-in images are shown in the 3$^{rd}$ and 6$^{th}$ rows. With the same radiation dose, obviously, the reconstructions with sinograms recommended by SAS achieve better results, where the reconstruction details are clearer than the others. Besides, we show the learned sampling trajectories of SAS across different cases and the dynamic sinogram selection results, which indeed contributes positive improvement to the final reconstruction. The display window is set to [-1000, 1000] in all cases.}
		\label{main-vis}
	\end{center}
\end{figure*}

\subsection{Patient-specific active reconstruction}
\label{patient-adaptive-recon}
\noindent\textbf{Quantitative comparison.} 
In this experiment, we quantitatively evaluate the effectiveness of our proposed SAS and GDS policies. The quantitative results of our two policies, as well as US and RS, are reported in Table~\ref{active-ablation}. Reconstructions with SAS and GDS achieve better performances across different settings except when ${k}_{max}$=$90$, where GDS policy is a little worse or comparable with the US since the restricted searching space limits its various hyperparameter selection (\textit{i.e.,} the performances).
Especially, the improvements over the US, which is commonly used in various algorithms, become enlarged when decreasing sinograms. This is caused by the increasing ``sinogram choosing freedom". In other words, when decreasing ${k}_{max}$ from $90$ to $15$, the whole selection range (0, ${\alpha}_{max}$) is fixed and the selection rate is reduced to 1/6. Therefore, we have more candidate angle positions in each selection. This highlights the advantage of the trajectory optimization of SAS and GDS in such extremely sparse scenarios. As clinical setup is different across hospitals, the model robustness to the involved photo noises is therefore important, and we test these policies with noise levels $L_1$ (Poisson noise level $5e^5$) and $L_2$ (Poisson noise level $1e^5$). Across all these noise levels, our SAS and GDS achieve consistently better reconstructions, which confirms its feasibility in practical deployment.

\noindent\textbf{Qualitative comparison.} To further understand the interaction between the adaptiveness of learning-based policies and the final reconstruction, we visualize the reconstructions of RS, US, and SAS in Fig.~\ref{main-vis}. The first three rows are the difference images, reconstructed images, and zoom-in images when ${k}_{max}=30$, respectively; the last three rows are when ${k}_{max}=15$. In accord with the above quantitative results, the reconstructed images with SAS show better performances in both intrathoracic tissues and bones. With such extremely sparse views, projection information is limited, and the details in intrathoracic parts are difficult to be recovered with both RS and US as shown in bottom right part of Fig.~\ref{main-vis}. While simultaneously optimizing sampling trajectories provides a potential to focus on the projections that are the most important for the final reconstructions. Besides, comparing the four images representing very different tissue distributions, the dynamically chosen sinograms are indeed patient-specific, caused by the different interactions between the reconstruction and each sampling sinogram. As the quantitative and qualitative performances are competitive, we only experiment with SAS policy later for simplicity.

\subsection{Active sampling process analysis}
\begin{figure}[t]
	\begin{center}
		\includegraphics[width=0.49\textwidth]{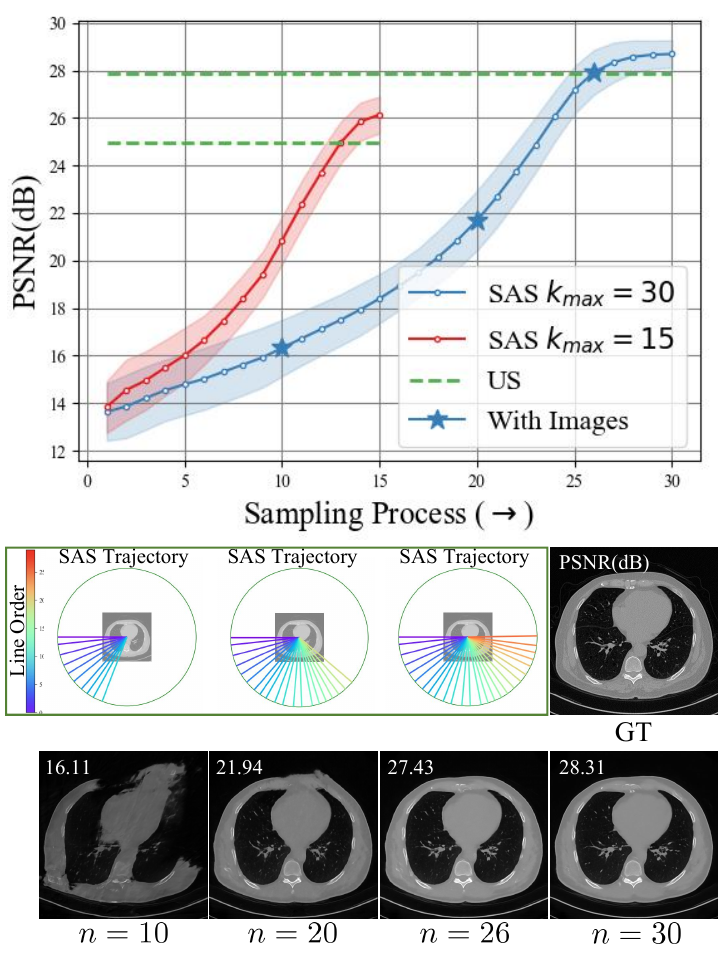}
	\end{center}
	\caption{We visualize the sampling process and intermediate reconstructed images of SAS when $n = 10, 20, 26$ (marked by `*' symbols on the SAS curve), and their corresponding learned trajectories.}
	\label{reduced-views}
\end{figure}

Then, we further visualize the whole sampling process when ${k}_{max}=15, 30$ in Fig.~\ref{reduced-views} 
As shown with the red and blue lines, the reconstruction performance is improved faster in the beginning, while suffering instability across samples. Then, with sufficient sampled views, the improvement slows down when sampling the last 13.5\% views, while model deviation on samples is smaller in this stage. 

\begin{table}[!t]
	\setlength\tabcolsep{3pt}
	\caption{We test the trained model when $k_{max} = 30$ on other scenarios, and show several quantitative results. Across different cases when $15<k_{max}<60$, our SAS policy exhibits better results than the US policy.}
	\label{robustness-across-different-cases}
	\begin{center}
		\resizebox{0.5\textwidth}{!}{
			\begin{tabular}{  l | lll | lll  }
				\hline
				\multirow{2}{*}{{NIH-AAPM}}
				& \multicolumn{3}{c|}{{US}}  &\multicolumn{3}{c}{{SAS (ours)}}      \\ 
				\cline{2-7}
				& PSNR(dB)&SSIM&RMSE&PSNR(dB)&SSIM&RMSE \\ \hline
				$k_{max}$ = 15 & 17.97$^{\pm1.11}$&.715$^{\pm.025}$&.127$^{\pm.014}$& \textbf{20.29$^{\pm1.49}$}& \textbf{.760$^{\pm.026}$}&\textbf{.098$^{\pm.012}$} \\
				$k_{max}$ = 30 & 27.87$^{\pm0.59}$&.845$^{\pm.017}$&.041$^{\pm.003}$& \textbf{28.69$^{\pm0.57}$} &\textbf{.855$^{\pm.017}$}&\textbf{.037$^{\pm.003}$}\\
				$k_{max}$ = 45 & 26.24$^{\pm1.53}$&.850$^{\pm.015}$&.050$^{\pm.011}$& \textbf{27.20$^{\pm1.56}$}& \textbf{.858$^{\pm.015}$}&\textbf{.045$^{\pm.011}$}\\
				$k_{max}$ = 60 & 24.50$^{\pm1.47}$&.850$^{\pm.015}$&.061$^{\pm.013}$& \textbf{25.03$^{\pm1.44}$}&\textbf{.852$^{\pm.015}$} &\textbf{.057$^{\pm.012}$}\\
				\hline
			\end{tabular}
		}
	\end{center}
\end{table}

\begin{table*}[t!]
	\setlength\tabcolsep{3pt}
	\begin{center}
		\caption{We test the robustness of our SAS on training data scales, and SAS shows comparable stability with US policy before decreasing data to 20\%. With limited 20\% training data, the SAS policy still shows the best performances.}
		\label{robustness-across-data-scales}
		\begin{tabular}{  l | lll | lll | lll  }
			\hline
			\multirow{2}{*}{{NIH-AAPM}}
			&\multicolumn{3}{c|}{{RS}}   
			& \multicolumn{3}{c|}{{US}}  &\multicolumn{3}{c}{{SAS (ours)}}      \\ 
			\cline{2-10}
			& PSNR(dB)&SSIM&RMSE& PSNR(dB)&SSIM&RMSE&PSNR(dB)&SSIM&RMSE \\ \hline
			20\% Training Data &20.52$^{\pm1.59}$&.753$^{\pm.041}$&.096$^{\pm.018}$ &24.23$^{\pm1.57}$&.783$^{\pm.043}$&.063$^{\pm.012}$  &\textbf{24.77$^{\pm1.12}$}&\textbf{.804$^{\pm.041}$}&\textbf{.058$^{\pm.008}$} \\
			40\% Training Data &23.62$^{\pm1.17}$&.807$^{\pm.025}$&.067$^{\pm.010}$&26.52$^{\pm1.53}$&.841$^{\pm.021}$&.048$^{\pm.012}$ &\textbf{27.10$^{\pm1.30}$}&\textbf{.845$^{\pm.015}$}&\textbf{.045$^{\pm.008}$}\\
			60\% Training Data &24.44$^{\pm1.04}$&.816$^{\pm.018}$&.060$^{\pm.008}$ &27.75$^{\pm1.55}$&.856$^{\pm.012}$&.042$^{\pm.011}$ &\textbf{28.24$^{\pm0.84}$}&\textbf{.852$^{\pm.015}$}&\textbf{.039$^{\pm.004}$}\\
			80\% Training Data &25.08$^{\pm1.42}$&.826$^{\pm.033}$&.056$^{\pm.012}$ &27.61$^{\pm1.17}$&.846$^{\pm.015}$&.042$^{\pm.007}$ &\textbf{28.44$^{\pm0.78}$}&\textbf{.854$^{\pm.015}$}&\textbf{.038$^{\pm.004}$}\\
			100\% Training Data&25.85$^{\pm1.16}$&.836$^{\pm.015}$&.053$^{\pm.008}$&27.87$^{\pm0.59}$&.845$^{\pm.017}$&.041$^{\pm.003}$& \textbf{28.69$^{\pm0.57}$} &\textbf{.855$^{\pm.017}$}&\textbf{.037$^{\pm.003}$}\\
			\hline
		\end{tabular}
	\end{center}
\end{table*}

\begin{table*}[]
	\setlength\tabcolsep{3pt}
	\begin{center}
		\caption{Testing results on different anatomies, including chest, brain, and abdomen, when $k_{max} = 30$. Our SAS policy achieves consistently better performances on all the anatomies, implying the robustness of the proposed sampling policy.}
		\label{different-pathological-parts}
		\begin{tabular}{  l | lll | lll | lll }
			\hline
			\multirow{2}{*}{{NIH-AAPM}}
			& \multicolumn{3}{c|}{{ Chest }}  
			&\multicolumn{3}{c|}{{ Brain }}
			&\multicolumn{3}{c}{{ Abdomen }}    \\ 
			\cline{2-10}
			& PSNR(dB)&SSIM&RMSE& PSNR(dB)&SSIM&RMSE&PSNR(dB)&SSIM&RMSE  \\ \hline
			RS &25.85$^{\pm1.16}$&.836$^{\pm.015}$&.053$^{\pm.008}$&29.03$^{\pm2.10}$&.908$^{\pm.032}$&.036$^{\pm.009}$&29.67$^{\pm2.29}$&.947$^{\pm.016}$&.034$^{\pm.011}$\\
			US  &27.87$^{\pm0.59}$&.845$^{\pm.017}$&.041$^{\pm.003}$& 33.33$^{\pm2.89}$&.942$^{\pm.025}$&.023$^{\pm.008}$ &31.69$^{\pm1.67}$&.952$^{\pm.011}$&.027$^{\pm.006}$  \\ 
			SAS (ours)  &\textbf{28.69$^{\pm0.57}$} &\textbf{.855$^{\pm.017}$}&\textbf{.037$^{\pm.003}$} &\textbf{33.74$^{\pm1.89}$}&\textbf{.954$^{\pm.010}$}&\textbf{.021$^{\pm.005}$}&\textbf{33.12$^{\pm1.70}$}&\textbf{.955$^{\pm.012}$}&\textbf{.022$^{\pm.005}$}  \\ 
			\hline
		\end{tabular}
	\end{center}
\end{table*}

Besides, we mark the US as the baseline with green dotted lines. With such comparison, we observe that our SAS achieves comparable performances when $n=13, 26$ in the two scenarios, respectively. This means that the sampling trajectory optimization of SAS reduces about 13.3\% views, \textit{i.e.,} 13.3\% radiation dose. The characteristic is especially important for protecting patient safety, and ensures that dynamically changing sampling trajectories according to the patient body representation is necessary.
Next, we show three reconstructed images when $n=10, 20, 26$, and their corresponding learned trajectories. The dynamical process shows that the sampling sparsity is still important as in the US, which helps cover body information as much as possible. Then, the adaptively-changing trajectories guide the model to focus on the parts which contribute most to the final reconstruction.


\subsection{Generalization over unseen ${k}_{max}$}
With the trained model, a practically deployed reconstruction model needs to be capable of dealing with unobserved settings, \textit{i.e.}, different ${k}_{max}$. Thus, we test our SAS trained with ${k}_{max}=30$ on other settings. As in Table~\ref{robustness-across-different-cases}, the performances across scenarios when ${k}_{max} = 15, 45, 60$ show better stability, compared with US. The improved stability is benefited from the modeling between sinograms and the final reconstruction. Especially, after training with the self-supervision of Eq.~(\ref{loss-IA}), our $\textbf{A}$ module can discriminate how the current candidate sinograms correlate with the image quality.


\subsection{Robustness on training data scale}
In medical image analysis, the limited data scale inhibits the research progress. Recently proposed Transformer-based methods~\cite{liu2021swin} in particular need large scale training data for good performances. Therefore, we test whether SAS suffers the problem, and compare results in Table~\ref{robustness-across-data-scales}. The three sampling policies all show stable performances with over 20\% training data, and our SAS keeps the similar improvement over RS and US. Then, with further decreasing training data to 20\%, all the three policies drops, which is caused by the too-limited training data, and the large distribution gap between training and testing data makes it hard to keep satisfactory performance. In such extremely difficult cases, our SAS still keeps a similar improvement, which probably benefits from the pretraining of \textbf{A} module.

\begin{figure*}[!t]
	\begin{center}
		\includegraphics[width=\textwidth]{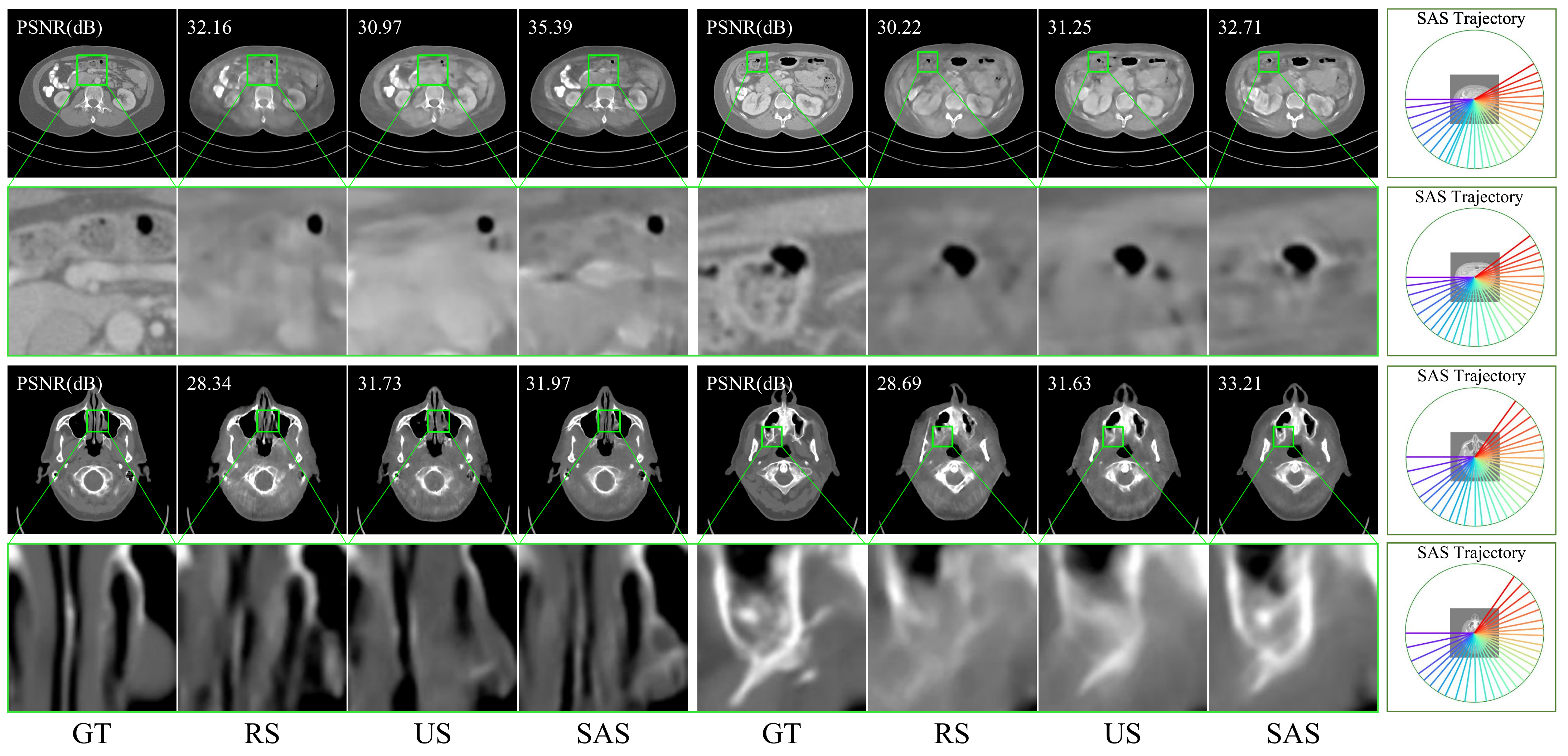}		
		\caption{We visualize the reconstructions of different anatomies when ${k}_{max}= 30$ in the 1$^{st}$ and 3$^{rd}$ rows, corresponding to abdomen and brain, respectively. The 2$^{nd}$ and 4$^{th}$ rows show the zoom-in images of the clinical RoI, and the learned trajectories are shown in the last column. Obviously, with the same radiation dose, learning-based SAS outputs better results, where the reconstruction details are clearer than the others. Besides, we exhibit the learned sampling trajectories across the four cases to show the dynamic sinogram selection results, which indeed contribute positive improvement to the final reconstruction.}
		\label{main-vis-pathological}
	\end{center}
\end{figure*}

\subsection{Robustness on anatomical structures}
\label{pathology}
As CT-based clinical decision making concerns multiple anatomical structures, such as the brain and abdomen, we conduct experiment based on these two structures in NIH-AAPM dataset to verify the robustness of our SAS policy. Specifically, we select 10 patients with abdomen CT images and 12 patients with brain CT images for training and use additional 4 patients with abdomen CT images and 3 patients with brain CT images for testing. The ${k}_{max}$ is set to 30 and the results are shown in Table.~\ref{different-pathological-parts}. We observe a consistent improvement of our SAS policy over RS and US. Especially, for the abdomen images, our policy obtains the 1.43 dB improvement over the US. For considerable analysis, we also visualize the images in Fig.~\ref{main-vis-pathological}, where our SAS reconstructions provide better clinical RoI recovery in zoom-in images. The accompanying learned trajectories are shown in the last column. 

\begin{table}[t!]
	\setlength\tabcolsep{3pt}
	\begin{center}
		\caption{Testing performance comparison on different backbones ($k_{max}$ = 30). Results show that our SAS policy is superior on both CNN and Transformer backbones}
		\label{different-backbones}
		\resizebox{0.5\textwidth}{!}{
			\begin{tabular}{  l | lll | lll  }
				\hline
				\multirow{2}{*}{{NIH-AAPM}}
				&\multicolumn{3}{c|}{{ US }}
				&\multicolumn{3}{c}{{ SAS }}    \\ 
				\cline{2-7}
				& PSNR(dB)&SSIM&RMSE&PSNR(dB)&SSIM&RMSE  \\ \hline
				CNN &27.87$^{\pm.59}$&.845$^{\pm.017}$&.041$^{\pm.003}$&\textbf{28.69$^{\pm.57}$} &\textbf{.855$^{\pm.017}$}&\textbf{.037$^{\pm.003}$}\\ 
				Transformer &28.10$^{\pm.81}$&.850$^{\pm.018}$&.039$^{\pm.004}$&\textbf{28.29$^{\pm.63}$}&\textbf{.853$^{\pm.018}$}&\textbf{.039$^{\pm.003}$}   \\ 
				\hline
			\end{tabular}
		}
	\end{center}
\end{table}

\subsection{Robustness on different backbones}
As a general framework for simultaneously sampling and reconstruction, we next substitute $\textbf{R}$ with recently proposed Transformer-based reconstruction architecture~\cite{wang2022dudotrans}, and compare the performances with previous CNN-based performances. 
As shown in Table~\ref{different-backbones}, compared with uniform sampling, our SAS achieves better performance, regardless of the network architecture. However, the performance improvement of SAS over US is bigger when the CNN background is utilized than when the Transformer is used, which indicates that SAS better matches with CNN-based architecture. In future, we will further explore a better combination of our SAS and other reconstruction frameworks, such as Transformer-based and deep-unfolding-based methods.

\begin{figure}[!t]
	\begin{center}
		\begin{minipage}[t]{0.24\textwidth}
			\begin{center}
				\includegraphics[width=\textwidth]{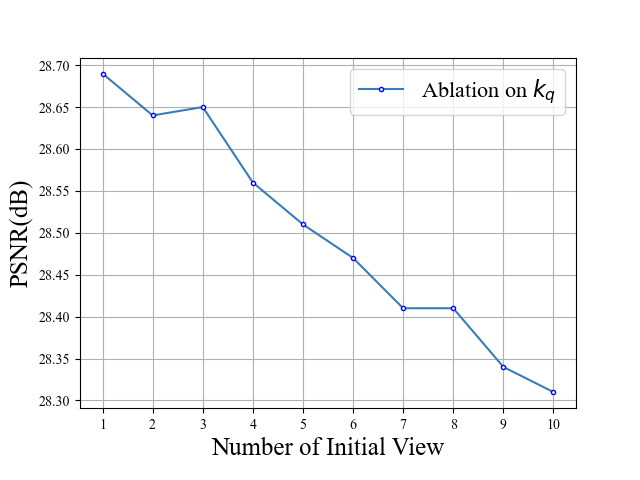}
				(a) Ablation on ${k}_{0}^{'}$
				\includegraphics[width=\textwidth]{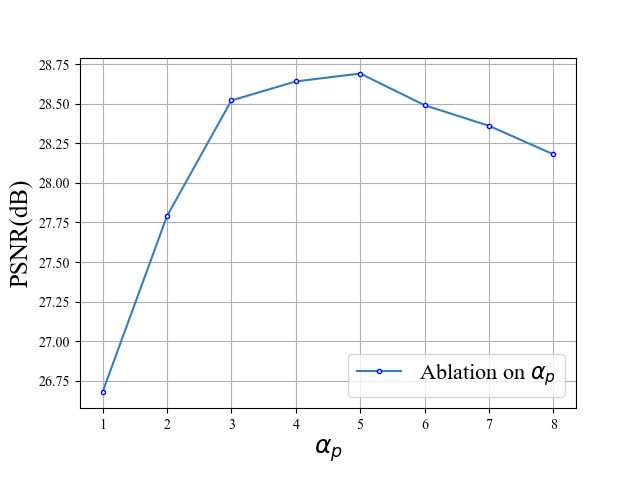}
				(c) Ablation on ${\alpha}_{p}^{'}$
			\end{center}
		\end{minipage}
		\begin{minipage}[t]{0.24\textwidth}
			\begin{center}
				\includegraphics[width=\textwidth]{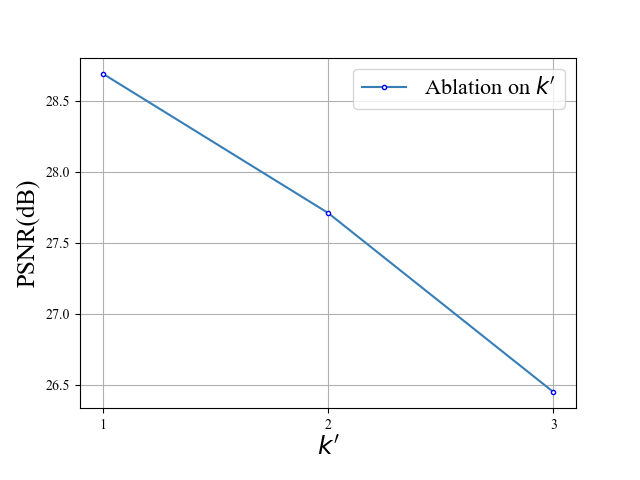}
				(b) Ablation on ${k}^{'}$
				\includegraphics[width=\textwidth]{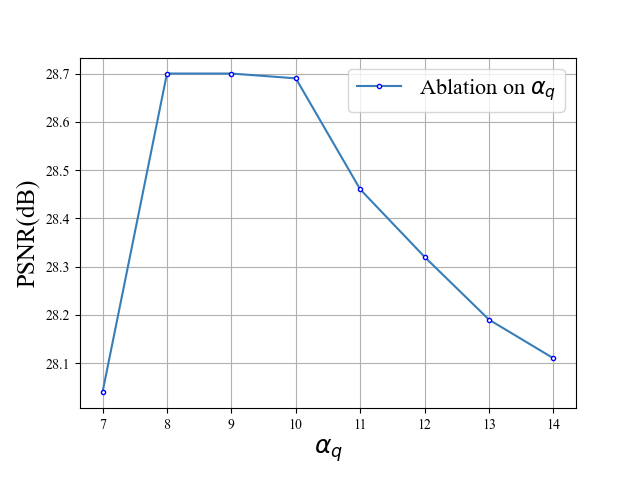}
				(d) Ablation on ${\alpha}_{q}^{'}$
			\end{center}
		\end{minipage}
		\caption{The conducted ablation experiments on the hyperparameters when reconstructing with SAS.}
		\label{ablation-hypers}
	\end{center}
\end{figure}

\begin{table*}[h!]
	\setlength\tabcolsep{3pt}
	\begin{center}
		\caption{Testing performance comparison on the VerSe datasets ($k_{max}$ = 30). The additional weighted loss indeed improves SAS policy performance with a larger margin over the US policy, verifying the impact of the loss on the view choice of SAS policy.}
		\label{roi-orientied}
		\begin{tabular}{  l | lll | lll  }
			\hline
			\multirow{2}{*}{{VerSe-Spine}}
			&\multicolumn{3}{c|}{{ Whole Image }}
			&\multicolumn{3}{c}{{ Region of Interest (RoI) }}    \\ 
			\cline{2-7}
			& PSNR(dB)&SSIM&RMSE&PSNR(dB)&SSIM&RMSE \\ \hline 
			US &28.75$^{\pm2.15}$&.911$^{\pm.009}$&.038$^{\pm.010}$&26.13$^{\pm2.36}$&.893$^{\pm.019}$& .051$^{\pm.015}$  \\ 
			US\_RoI &28.82$^{\pm1.29}$&.904$^{\pm.015}$&.037$^{\pm.006}$& 26.47$^{\pm1.74}$&.894$^{\pm.022}$&.047$^{\pm.010}$  \\ 
			SAS (ours) &\textbf{29.24$^{\pm1.01}$}&\textbf{.919$^{\pm.009}$}&\textbf{.035$^{\pm.004}$}&\textbf{27.83$^{\pm1.93}$}&\textbf{.917$^{\pm.017}$}&\textbf{.042$^{\pm.009}$} \\ 
			SAS\_RoI (ours)&\textbf{29.91$^{\pm1.00}$}&\textbf{.913$^{\pm.009}$}&\textbf{.032$^{\pm.004}$}&\textbf{28.86$^{\pm1.41}$}&\textbf{.916$^{\pm.016}$}&\textbf{.037$^{\pm.006}$} \\ 
			\hline
		\end{tabular}
	\end{center}
\end{table*}

\subsection{Ablation on hyperparameters choices}
We then discuss the choice of contained hyperparameters in terms of chest CT image reconstruction when $k_{max}=30$. The conducted ablation experiments of four hyperparameters \{${k}_{0}^{'}$,  ${k}^{'}$,  ${\alpha}_{p}^{'}$, ${\alpha}_{q}^{'}$\} used in testing is exhibited in Fig.~\ref{ablation-hypers}. In subfigure (a), when we fix the other hyperparameters and gradually increase ${k}_{0}^{'}$ from 1 to 10, the reconstruction performance decreases step by step. Note that when increasing ${k}_{0}^{'}$, the searching freedom is gradually tightened which limits the final reconstructions. Then we fix other hyperparameters and explore the choice of ${k}^{'}$ as in subfigure (b). The best performance is achieved when ${k}^{'}$ equals 1. When increasing ${k}^{'}$ to 2 and 3, reconstruction performance decreases 0.98 and 2.24 dB since we set $k$ as 1 in training. When further increasing $k$ in training, the decrease is alleviated. However, a larger $k$ comes with larger selection freedom and hence raises the training difficulty; thus we set $k$ and ${k}^{'}$ to 1 in all our previous experiments. Next, with a similar procedure, we test the choice of ${\alpha}_{p}^{'}$ and ${\alpha}_{q}^{'}$ and show the results in subfigures (c) and (d). In such way, we obtain optimal choice when ${\alpha}_{p}^{'}$ and ${\alpha}_{q}^{'}$ equal to 5 and 10, respectively. With the above described manner, we have achieved the best reconstruction performance as Table~\ref{active-ablation} reports.

\subsection{RoI-aware reconstruction}
\begin{figure}[!t]
	\begin{center}
		\includegraphics[width=0.5\textwidth]{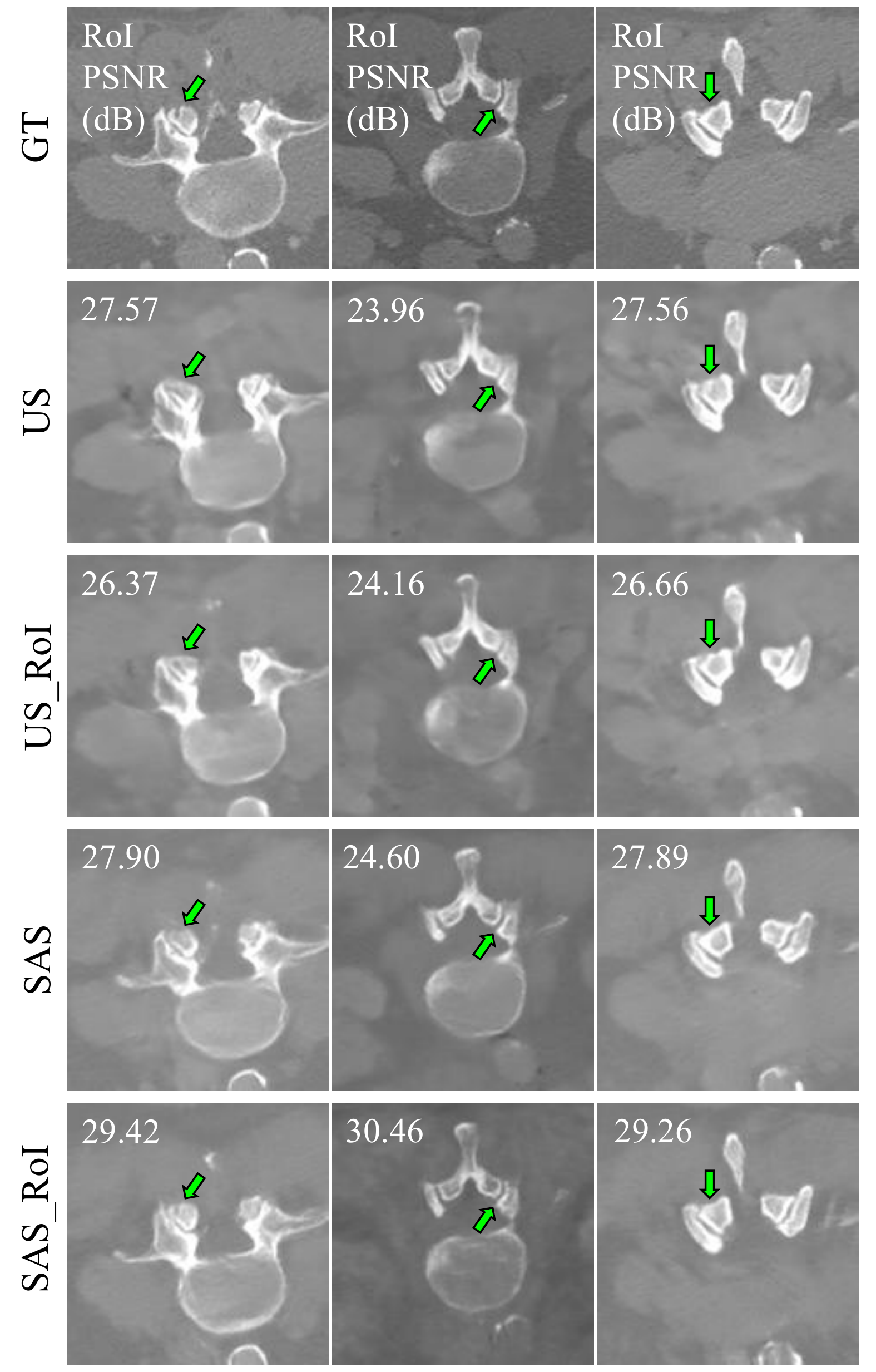}		
		\caption{We visualize the RoI(spine in our experiments) reconstructions of US, US\_RoI, SAS, and SAS\_RoI. Results across different slices show that SAS\_RoI with the weighted loss provides the best recovery, even if the SAS has recovered most details. The display window is set to [-1000, 1000].}
		\label{roi}
	\end{center}
\end{figure}

With the active sampling procedure, a direct potential application is a combination with downstream tasks since they provide a prior clinically-concerned RoI. Targeting the exploration of the interaction between the additional weight loss introduced in Eq.~(\ref{roi-loss}) and our SAS policy, we investigate RoI-aware reconstruction and conduct experiments on VerSe~\cite{SEKUBOYINA2021102166} covering the spine. Specifically, we employ partial slices of seven patients covering spines, of which five are for training and the others for testing. To verify the influence, we equip the weighted loss for both US and SAS policies, named US\_RoI and SAS\_RoI, and report the quantitative results in Table~\ref{roi-orientied}. Comparing US and SAS, our policy achieves better reconstruction of both the whole image and RoI. Further, when additionally employing Eq.~(\ref{roi-loss}), SAS\_RoI increases about 1dB on the region containing spines, even if the SAS has achieved 27.83 dB. The improvements on the RoI ensure that the active sampling process is guided by the weighted reconstruction loss. Besides, we visualize the RoI image in Fig.~\ref{roi}, including three cases with different spine parts. Coinciding with the quantitative results, the US is almost not affected by the weighted RoI loss. In contrast, reconstructions of SAS\_RoI provide clearer recovery on the clinical concerned region, which is valuable for practical diagnosis.

\section{Discussions}
\noindent\textbf{Comparisons with greedy sampling.} To further explore the upper bound of our learning-based sampling policies, we test a training-free greedy sampling method. Assuming that the sampled sinogram positions are in a set $\cal{S}$. Specifically, we initialize $\cal{S}$$= \{ 0^{\circ} \}$ with US, and the left positions set $\cal{\overline{S}}= \{$$1^{\circ}$,$ 2^{\circ},$$ \ldots, $$359^{\circ}\}$ are candidates for the next sampling. Then, we reconstruct with $\cal{S}$ + $\overline{s}$, where  $\overline{s}$ $\in \overline{S}$. Choosing the $\overline{s}$ providing the largest reconstruction improvement, based on PSNR, over the reconstruction with $\cal{S}$ only, and denote it as $\overline{s}'$. We next update $\cal{S}$=$\cal{S}\bigcup$$\overline{s}'$. The sampling process continues till $| \cal{S} | $$= {k}_{max}$. To be fair, we use our trained reconstructor here since our reconstructor is more robust to ${k}_{max}$ than the US-trained model. We name such training-free greedy sampling as GS and compare it with SAS and US in Table.~\ref{discussion-greedy}. Since the greedy choosing process is too computationally expensive, we conduct only two cases when $k_{max}$ = 15 and 30. The obvious improvement is achieved compared with the US in both cases. However, our SAS still achieves improvement when $k_{max}$ = 30 over GS, which is because of the training benefit. Such comparisons verify that our method indeed closely approximates such a completely greedy sampling method. 

\noindent\textbf{Time comparison} Considering the real-time practical deployment, we compare the performance improvement versus computation time in Table~\ref{discussion-time}. Our SAS achieves similar results with GS, while the US performance is much lower, especially when ${k}_{max}$ is smaller. Simultaneously, the inference time of SAS is much shorter than that of GS, while about 7 and 19 times slower than the predefined US when ${k}_{max}$ equals 15 and 30. Balancing the performance improvement versus additionally introduced computation, SAS is practically feasible. 
Besides, when decreasing ${k}_{max}$ with the fixed ${\alpha}_{max}$, the accompanying larger sampling searching space brings larger improvement with our SAS. Furthermore, the smaller ${k}_{max}$ decreases the active search time, which enlarges the advantage of our SAS in extremely sparse scenarios. In contrast, the iterative computation time monotonically increases when increasing ${k}_{max}$. The smaller searching space also limits the performance of SAS and the completely greedy GS. With such comparison, we hypothesize that the sparser and larger sampling searching space would be beneficial to such learning-based sampling policies, and vice versa.

\begin{table}[t!]
	\setlength\tabcolsep{3pt}
	\begin{center}
		\caption{Comparisons with training-free reconstruction-benefited greedy sampling (GS) method when $k_{max}$ = 15, 30.}
		\label{discussion-greedy}
		\resizebox{0.5\textwidth}{!}{
			\begin{tabular}{  l | lll | lll  }
				\hline
				\multirow{2}{*}{{NIH-AAPM}}
				&\multicolumn{3}{c|}{{ $k_{max}=15$ }}
				&\multicolumn{3}{c}{{ $k_{max}=30$ }}    \\ 
				\cline{2-7}
				& PSNR(dB)&SSIM&RMSE&PSNR(dB)&SSIM&RMSE  \\ \hline
				US &24.98$^{\pm.59}$&.813$^{\pm.017}$&.057$^{\pm.004}$&27.87$^{\pm.59}$&.845$^{\pm.017}$&.041$^{\pm.003}$\\
				GS&\textbf{26.41}$^{\pm.74}$&.826$^{\pm.016}$&\textbf{.044}$^{\pm.004}$&28.57$^{\pm.64}$&.850$^{\pm.018}$&.037$^{\pm.003}$   \\ 
				SAS (ours)&26.16$^{\pm.80}$&\textbf{.828}$^{\pm.017}$&.049$^{\pm.005}$&\textbf{28.69$^{\pm.57}$} &\textbf{.855$^{\pm.017}$}&\textbf{.037$^{\pm.003}$}\\ 
				\hline
			\end{tabular}
		}
	\end{center}
\end{table}

\begin{table}[t!]
	\setlength\tabcolsep{3pt}
	\begin{center}
		\caption{Comparisons of PSNR and computation time among US, GS and SAS methods when $k_{max}$ = 15, 30.}
		\label{discussion-time}
		\begin{tabular}{  l | rr | rr  }
			\hline
			\multirow{2}{*}{{NIH-AAPM}}
			&\multicolumn{2}{c|}{{ $k_{max}=15$ }}
			&\multicolumn{2}{c}{{ $k_{max}=30$ }}    \\ 
			\cline{2-5}
			& PSNR(dB)&Time (s) &PSNR(dB)&Time (s)  \\ \hline
			US&24.98&\textbf{0.33} &27.87& \textbf{0.23}\\
			GS &\textbf{26.41}& 586.70&28.57&1187.55    \\ 
			SAS (ours) &26.16&2.42&\textbf{28.69}&4.34\\ 
			\hline
		\end{tabular}
	\end{center}
\end{table}

\noindent\textbf{Comparisons with the dynamically collimator-based RoI reconstruction.} Targeting different clinical tasks, the CT images need to focus on task-specific regions, i.e. RoI, providing enough information for clinicians in diagnosis. This also helps reduce quite a lot of radiation imposed on other diagnosis-irrelevant anatomies. For this purpose, a commonly practical used RoI CT reconstruction method is the dynamically collimated X-Ray beam~\cite{heuscher2011ct}, which utilizes the dynamic collimator to limit the exposure. This practical solution considers the RoI from the physical perspective and requires that the collimator velocity matches with the X-Ray source to exactly locate the RoI. While our SAS-based RoI reconstruction is from the algorithmic perspective, and the ideal \textbf{A} is capable to sample task-relevant sinograms. In this way, our algorithm is more suitable for before-diagnosis imaging, and the dynamically collimated X-Ray beam RoI reconstruction is more suitable for after-diagnosis imaging.

\noindent\underline{\textit{Further combination of both methods.}} Although SAS-based RoI reconstruction differs from the collimator-based solution, the two methods don't conflict with each other since they are from two parallel viewpoints. Further combining the algorithmic perspective of SAS and the physics-motivated collimator is a potentially clinical design.

\section{Conclusions}
In this work, we break through the limitation of uniform sampling in CT by proposing a novel framework of actively sampling sinograms that are adaptive to a patient for CT reconstruction with a reduced radiation dose. It is done by the specifically-designed \textbf{A} module for selecting the sinograms to acquire and \textbf{R} module for reconstructing CT images. In this way, the whole sampling trajectory takes into account the individual factors, and the later sampled sinogram position is determined by the pre-sampled ones. With such a design, we propose two sampling policies, called SAS and GDS, both achieving better reconstruction performances with the same reconstructor. Furthermore, when fusing the RoI information of clinically-concerned anatomy, the active sampling policy achieves even better reconstruction, especially in the RoI. However, the computational efficiency heavily depends on the introduced hyperparameter $k$, which we would explore to raise reconstruction efficiency in future.


%

\bibliographystyle{IEEEtran} 
\bibliography{general_ref}





\ifCLASSOPTIONcaptionsoff
  \newpage
\fi

\end{document}